\documentstyle[12pt,epsf]{article}
\begin{document}
\def \beq{\begin{equation}}
\def \eeq{\end{equation}}
\def \es{E$_{\rm 6}$}
\def \g{{\rm GeV}}
\def \seff{\sin^2 \theta_{\rm eff}}
\rightline{DOE/ER/40561-64-INT99}
\rightline{EFI-99-34}
\rightline{hep-ph/9907524}
\vspace{0.5in}
\centerline{\bf ATOMIC PARITY VIOLATION AND PRECISION}
\centerline{\bf ELECTROWEAK PHYSICS - AN UPDATED ANALYSIS\footnote{Submitted to
Physical Review D.}} 
\vspace{0.5in}
\centerline{\it Jonathan L. Rosner}
\centerline{\it Institute for Nuclear Theory}
\centerline{\it University of Washington, Seattle, WA 98195}
\bigskip
\centerline{and}
\bigskip
\centerline{\it Enrico Fermi Institute and Department of Physics}
\centerline{\it University of Chicago, Chicago, IL 60637
\footnote{Permanent address.}}
\bigskip


\centerline{\bf ABSTRACT}
\begin{quote}
A new analysis of parity violation in atomic cesium has led to the improved
value of the weak charge, $Q_W({\rm Cs}) = -72.06 \pm 0.46$.  The implications
of this result for constraining the Peskin-Takeuchi parameters $S$ and $T$ and
for guiding searches for new $Z$ bosons are discussed.
\end{quote}

\leftline{PACS Categories: 11.30.Er, 12.15.Ji, 12.15.Mm, 12.60.Cn}
\bigskip


One prediction of the unified theory of weak and electromagnetic interactions
\cite{GWS} is the existence of parity-violating effects in atoms. In the latest
contribution \cite{NewCs} to this subject through the study of such effects in
atomic cesium \cite{Cs88,Cs97}, the JILA/Boulder group has performed
measurements that reduce uncertainties in previous theoretical calculations of
atomic physics corrections \cite{Csth}. While there is no substitute for
carrying out such calculations to the requisite higher order in many-body
perturbation theory, it is worth examining the implications of the resulting
weak charge, $Q_W({\rm Cs}) = -72.06 \pm 0.28_{\rm expt} \pm 0.34_{\rm theor} 
= -72.06 \pm 0.46$, which represents a considerable improvement with respect to
previous values in this and other \cite{Bi,Pb,TlS,TlO} atoms. The present note
updates previous analyses \cite{MR,PT,APV95,APV97,TLG}, with special emphasis
on the role of the new measurement.  We indicate the effect of fits to
precision electroweak observables in which the new measurement is included or
omitted, and discuss the possibility \cite{MR,Cas} that a small discrepancy of
$Q_W({\rm Cs})$ with respect to electroweak predictions is due to the exchange
of a new neutral vector gauge boson $Z'$.  The weak charges $Q_W$ provide
unique information in such fits \cite{MR,PS,JRRC}. 

Data and theoretical expectations are presented in Table 1. The notation and
formalism are the same as in Refs.~\cite{APV95} and \cite{APV97}.  As mentioned
previously, we use a subset of the data in which the effects of correlations
are minimized, but which have the dominant statistical weight.  For fits to the
complete data set, see, e.g., \cite{EWWG} or \cite{EL}.  Some new features with
respect to our previous fits include the following: 

\begin{table}
\begin{center}
\caption{Electroweak observables described in fit}
\medskip
\begin{tabular}{c c c} \hline
Quantity      &   Experimental   &   Theoretical \\
              &      value       &    value      \\ \hline
$Q_W$ (Cs)    & $-72.06 \pm 0.46^{~a)} $  &  $ -73.19^{~b)} - 0.80S - 0.007T$\\
$Q_W$ (Tl)    & $-115.0 \pm 4.5^{~c)} $ &  $ -116.8^{~d)} -1.17S - 0.06T$ \\
$M_W~(\g/c^2)$ & $80.394 \pm 0.042^{~e)}$  & $80.315^{~f)} -0.29S + 0.45T$ \\
``$M_W$'' $(\g/c^2)$ & $80.36 \pm 0.21^{~g)}$ & $80.315^{~f)}
  -0.29S + 0.52T^{~h)}$ \\
``$M_W$'' $(\g/c^2)$ & $80.24 \pm 0.11^{~i)}$ & $80.315^{~f)}
  -0.54S + 0.70T^{~h)}$ \\
$\Gamma_{\ell\ell}(Z)$ (MeV) & $83.958 \pm 0.089^{~j)}$ & $83.92^{~f)} -0.18S
+ 0.78T$ \\
$\seff$ & $0.23195 \pm 0.00023^{~j)}$ & $0.23200^{~f)}
 + 0.0036S - 0.0026T$ \\
$\seff$ & $0.23099 \pm 0.00026^{~k)}$ & $0.23200^{~f)}
 + 0.0036S - 0.0026T$ \\ 
$m_t~(\g/c^2)$ & $174.3 \pm 5.1^{~l)}$ & $173.9 + 241S + 82T $ \\ \hline
\end{tabular}
\end{center}
\leftline{$^{a)}$ {\small Weak charge in cesium \cite{NewCs} incorporating 
recalculated atomic physics corrections}}
\leftline{$^{b)}$ {\small Calculation \cite{MR} incorporating electroweak
corrections, updated in \cite{TLG}}}
\leftline{$^{c)}$ {\small Weak charge in thallium \cite{TlS,TlO} incorporating 
atomic physics corrections \cite{Tlth}}}
\leftline{$^{d)}$ {\small Calculation incorporating electroweak
corrections \cite{PSBL}}}
\leftline{$^{e)}$ {\small Average of direct hadron collider and LEP II
measurements \cite{Wmass}}}
\leftline{$^{f)}$ {\small Calculation by \cite{TLG} based on results of the
program ZFITTER 4.9 \cite{ZF}}} 
\leftline{$^{g)}$ {\small CCFR value from deep inelastic neutrino scattering
 \cite{CCFR}}}
\leftline{\qquad{\small for $m_t = 173.9~\g/c^2$ and $M_H = 300~\g/c^2$}}
\leftline{$^{h)}$ {\small Approximate dependence including residual
corrections}}
\leftline{$^{i)}$ {\small NuTeV value from deep inelastic neutrino scattering
 \cite{NuTeV}}}
\leftline{\qquad{\small for $m_t = 173.9~\g/c^2$ and $M_H = 300~\g/c^2$}}
\leftline{$^{j)}$ {\small LEP average as of July, 1999 \cite{Mnich,Brau}}}
\leftline{$^{k)}$ {\small From left-right asymmetry and forward-backward
left-right asymmetry at SLD \cite{Brau}}}
\leftline{$^{l)}$ {\small See Ref.~\cite{mt}}}
\end{table}

\begin{enumerate}

\item We use a new, more precise value $\alpha^{-1}(M_Z) = 128.933 \pm 0.021$
\cite{alpha}. 

\item The nominal top quark mass is now taken to be $173.9~\g/c^2$; the nominal
Higgs mass continues to be $300~\g/c^2$.  This permits us to use the
calculations of Ref.~\cite{TLG} for several quantities, including $M_W$,
$\Gamma_{\ell\ell}(Z)$, and $\seff$. 

\item The fits are performed both with and without the new Cs data
\cite{NewCs}, in order to estimate their impact. 

\item The precision of the world average value of $M_W$ \cite{Wmass} has
improved considerably as a result of new measurements from LEP II and the
Fermilab Tevatron.

\item We take account of a new measurement of the neutral-current to
charged-current ratio in deep inelastic neutrino scattering \cite{NuTeV}. We
present the result of this measurement, as well as that of a previous one
\cite{CCFR}, in terms of an effective $W$ mass corrected for our nominal values
of $m_t$ and $m_H$. This correction amounts to $-0.02~\g/c^2$ for \cite{NuTeV}
and $+ 0.01~\g/c^2$ for \cite{CCFR}.  The $S$ and $T$ coefficients
differ from those in $M_W$ since NuTeV measures the Paschos-Wolfenstein
\cite{PW} ratio $R_- \equiv [\sigma_{NC}(\nu N) - \sigma_{NC}(\bar \nu
N)]/[[\sigma_{CC}(\nu N) - \sigma_{CC}(\bar \nu N)]$, while CCFR measures
essentially $R_\nu \equiv \sigma_{NC}(\nu N)/\sigma_{CC}(\nu N)$. 

\item The precision of the LEP I values for $\Gamma_{\ell\ell}(Z)$ and $\seff$
\cite{Mnich}, the SLD value of $\seff$ \cite{Brau}, and the top quark mass
measurement \cite{mt} continues to improve. In our analysis we have combined
the values of $\seff$ from LEP I and SLD, with a scale factor \cite{PDG} of
$\sqrt{\chi^2} = 2.77$, and added in quadrature an error on the predicted value
of $\pm 0.00009$ due to the error in $\alpha(M_Z)$, to obtain a value $\seff =
0.23153 \pm 0.00048$ used as a single input to the fit.  We include values
of $\seff$ obtained at LEP both with purely leptonic asymmetries and with the
help of quark asymmetries such as $A_{FB}^b$, assuming them to be governed by
the predictions of the standard model.  The degree to which this fails to
be true \cite{Brau}, for example as a result of non-standard $b$ quark
couplings to the $Z$, is an interesting possibility not considered here.  
The LEP values of $\seff$ obtained from purely leptonic asymmetries do appear
to be more consistent with the SLD value.

\end{enumerate}

The results are shown in Figs.~1 and 2.  In Fig.~1 we have not imposed the
constraint of the top quark mass, while in Fig.~2 this constraint has been
included.

\begin{figure}
\centerline{\epsfysize = 3 in \epsffile {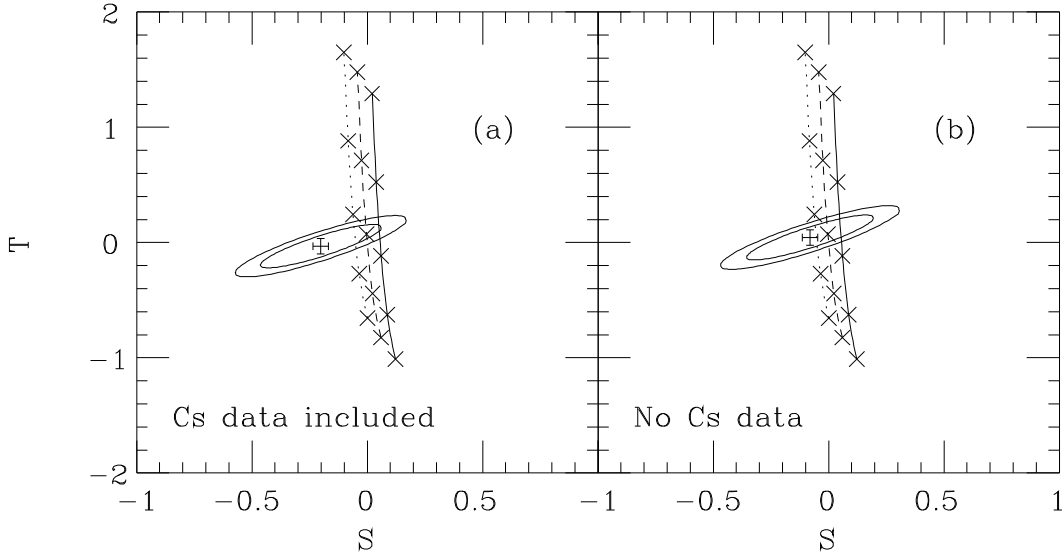}}
\caption{Allowed ranges of $S$ and $T$ at 68\% (inner ellipses) and 90\% (outer
ellipses) confidence levels, corresponding to $\chi^2 = 2.3$ and 4.6 above the
minima (crosses at center of ellipses).  Dotted, dashed, and solid lines
correspond to standard model predictions for $M_H = 100$, 300, 1000 GeV/$c^2$.
Symbols $\times$, from bottom to top, denote predictions for $m_t = 100$, 140,
180, 220, and 260 GeV/$c^2$. (a) Fit including APV experiments with present
errors; (b) fit excluding new Cs measurement.}
\end{figure}

The central values $S_0$ and $T_0$ implied by each of the fits are summarized
in Table 2.  We do not fit separately for the Peskin-Takeuchi parameter $U$,
but set it equal to zero.  A fit to similar data without the addition of
the new Cs results finds \cite{TLG} $S = -0.30 \pm 0.13$, $T = -0.14 \pm
0.15$, $U = 0.15 \pm 0.21$.

\begin{table}
\caption{Central values of $S$ and $T$ implied by fits to electroweak data,
omitting new Cs data, $m_t$ value, or both.}
\begin{center}
\begin{tabular}{c c c c} \hline \hline
Data omitted &  $S_0$   &  $T_0$  & Predicted $Q_W({\rm Cs})$ \\ \hline
$m_t$        & $-0.20$  & $-0.03$ & $-73.03$ \\
$m_t$ and Cs & $-0.08$  & $0.04 $ & $-73.13$ \\
None         & $-0.029$ & $0.083$ & $-73.17$ \\
Cs           & $-0.026$ & $0.080$ & $-73.17$ \\ \hline
\end{tabular}
\end{center}
\end{table}

In the absence of the $m_t$ constraint (Fig.~1), the new Cs analysis leads to a
small shift of the overall fit away from predictions of the standard
electroweak theory for the minimum acceptable Higgs boson mass (roughly
$95~\g/c^2$ \cite{MH}).  The change in the central value of the parameter $S$
is $-0.12$.  In the presence of the $m_t$ constraint (Fig.~2), the fit is
affected only very slightly by the Cs result.  The observed value of $Q_W$ then
differs from the predicted value by 2.4 standard deviations. Strictly speaking,
we should have omitted the Tl results from the fits when omitting Cs.  However,
their impact is much smaller than that of Cs. 

\begin{figure}
\centerline{\epsfysize = 3 in \epsffile {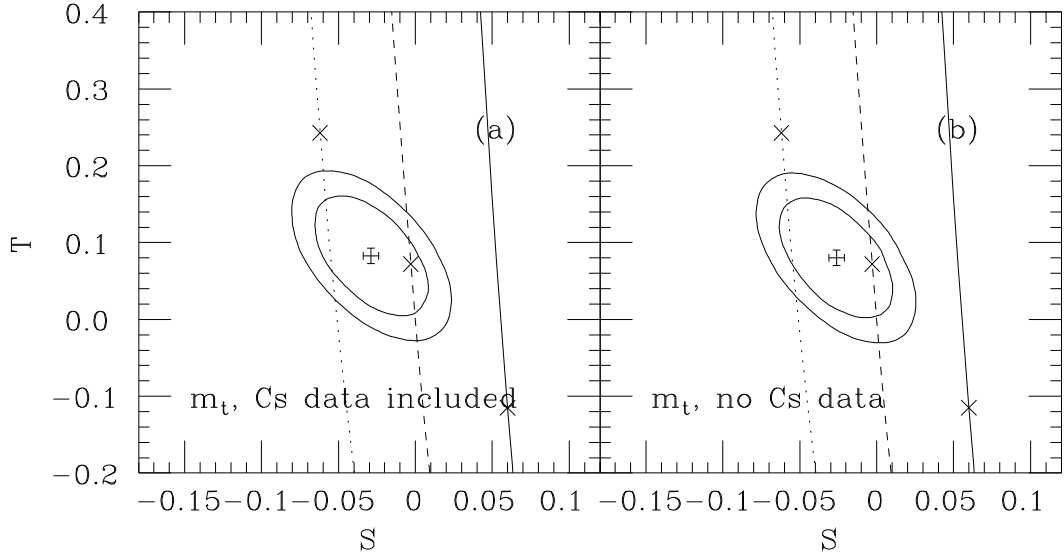}}
\caption{Magnified view of Figure 1.  Dotted, dashed, and solid lines
correspond to standard model predictions for $M_H = 100,~300,~1000~\g/c^2$.
Symbols $\times$ denote predictions for $m_t = 180~\g/c^2$ on each curve. The
constraint $m_t = 173.8 \pm 5~\g/c^2$ has been imposed. (a)  New Cs value
\protect \cite{NewCs} included; (b) New Cs value omitted.} 
\end{figure}

We now explore the implications of the small discrepancy between the observed
and predicted values of $Q_W({\rm Cs})$ in terms of an extra $Z$, as suggested
in Refs.~\cite{MR} and \cite{APV95}.  Our results differ slightly from those
of Ref.~\cite{Cas} as a consequence of a different standard-model prediction
for $Q_W$.

We consider a $Z'$ which is a linear combination of the $Z_\chi$ and $Z_\psi$ 
\cite{LRR}, two neutral bosons which arise in \es~theories: $Z' = Z_\psi \cos
\phi + Z_\chi \sin \phi$. Here $\phi$ is the angle called $\theta$ in
Ref.~\cite{LR}.  The $Z_\psi$ is the gauge boson associated with the symmetry
U(1)$_\psi$ when \es~breaks down to SO(10) $\times$ U(1)$_\psi$; the $Z_\chi$
is the gauge boson associated with the symmetry U(1)$_\chi$ when SO(10) breaks
down to SU(5) $\times$ U(1)$_\chi$. The change in $Q_W$ at tree level due to an
unmixed $Z'$ is then \cite{APV95} 
$$
\Delta Q^{\rm new}_{W~{\rm tree}} \simeq 0.4 (2N + Z) (M_W/M_{Z'})^2 f(\phi)
~~~,
$$
\beq
f(\phi) \equiv \sin \phi [\sin \phi - (5/3)^{1/2} \cos \phi]~~~.
\eeq
In order to fit the positive value of $\Delta Q^{\rm new}_{W~{\rm tree}} = 1.10
\pm 0.46$, we need $\phi$ to lie between $\tan^{-1}(5/3)^{1/2} = 52.2^{\circ}$
and $180^\circ$.  The corresponding values of $M_{Z'}$ leading to such a
contribution are shown for the central value and $\pm 1 \sigma$ limits on $Q_W$
by the curves in Fig.~3.  Typical direct lower limits from the CDF
Collaboration on masses of a $Z'$ depend to some extent on $\phi$, but lie
around $600~\g/c^2$ \cite{EL,CDFZp}.  At the $1 \sigma$ level, one can thus
account for the discrepancy between the observed and predicted values of
$Q_W({\rm Cs})$ for values of $\phi$ between about $70^\circ$ and $160^\circ$.
This includes the values $\phi = 90^\circ$ ($Z' = Z_\chi$) and $\phi =
127.8^\circ$ ($Z' = Z_{\rm I}$, where the subscript denotes an ``inert'' SU(2)
subgroup of \es~\cite{LRR,E6} in the decomposition \es$ \to$ SU(6) $\otimes$
SU(2)$_{\rm I}$.) 

To conclude, reanalysis of an atomic parity violation experiment in Cs
\cite{NewCs} affects fits of electroweak parameters to a small but perceptible
degree, when information on the top quark mass is not included.  When this
information is added, however, the fits are nearly independent of the Cs
result, which differs from the standard model prediction by 2.4 standard
deviations.  This difference can be reproduced by the inclusion of a new $Z'$,
lying above present experimental limits of about $600~\g/c^2$ in mass, for a
range of the parameter $70^\circ \le \phi \le 160^\circ$ characterizing the new
boson.  If it exists at a mass accessible to Run II of the Fermilab Tevatron,
this boson must be very weakly mixed with the standard $Z$ in order to avoid a
number of constraints associated with precision electroweak observables
\cite{EL}. 

Despite the consistency of the new measurements in Cs with more precisely
specified matrix elements \cite{NewCs}, a calculation of atomic physics effects
in Cs whose accuracy matches that of the experimental measurement is sorely
needed.  The last such calculations \cite{Csth} need to be extended to higher
order in many-body perturbation theory to confirm the optimism inherent in the
small theoretical error quoted in Ref.~\cite{NewCs}.  An improved determination
of the neutron charge radius in Cs also would be helpful, since present
uncertainty in this quantity may constitute an error at least as large as that
($\Delta Q_W \simeq 0.1$) associated with electroweak radiative corrections
\cite{CV,SP}. There is room for considerable improvement in the overall error
on $Q_W({\rm Cs})$ if this program proves successful. 

I am indebted to J. F. Beacom, E. N. Fortson, J. Sapirstein, and C. E. Wieman
for useful discussions. I wish to thank the Institute for Nuclear Theory at the
University of Washington and the Fermilab Theory Group for hospitality during
this work, which was supported in part by the United States Department of
Energy under Grant No. DE FG02 90ER40560. 

\begin{figure}
\centerline{\epsfysize = 5.5 in \epsffile {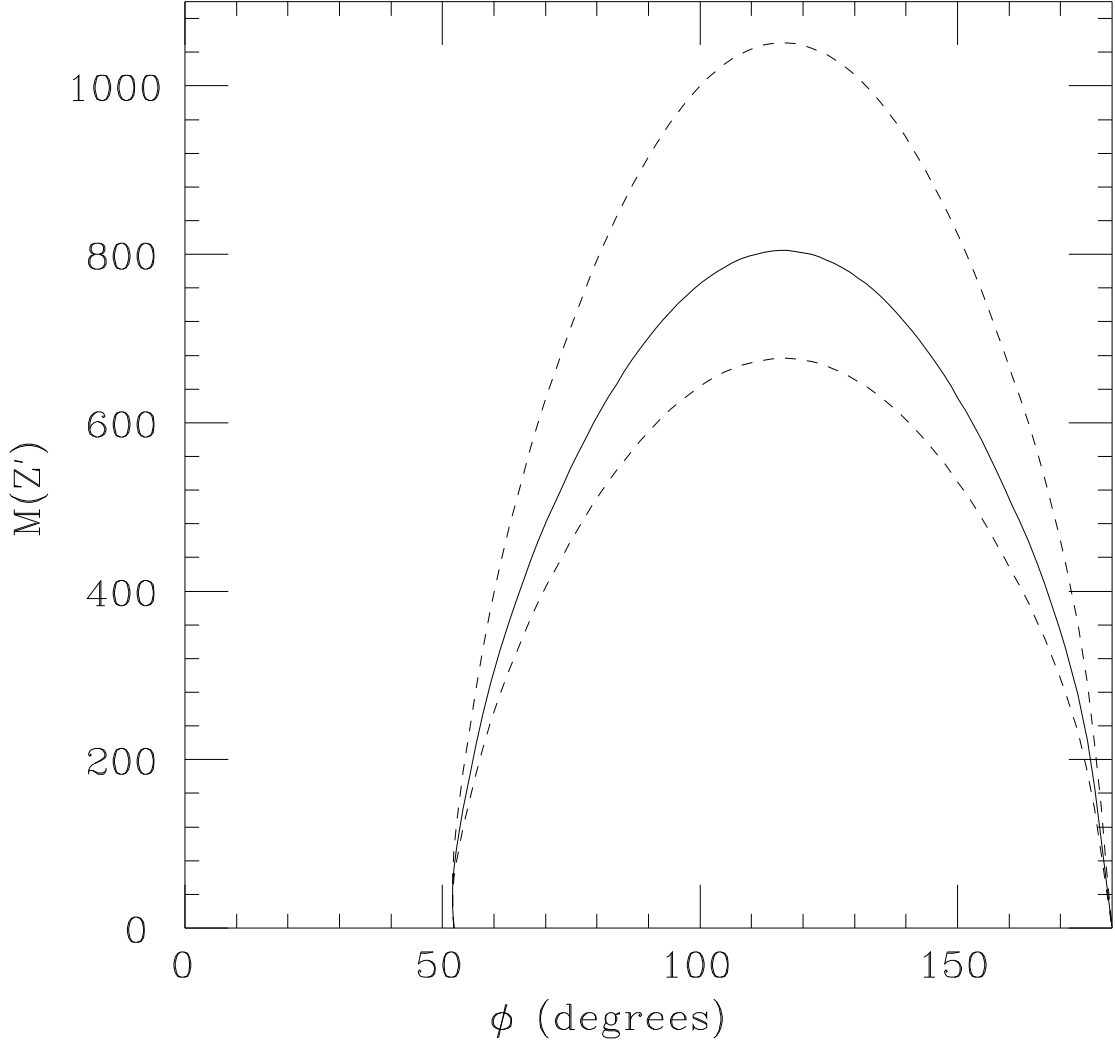}}
\caption{Values of $M(Z')$ corresponding to central value (solid line)
and $\pm 1 \sigma$ errors (dashed lines) of $Q_W({\rm Cs})$ in a model where
the discrepancy with respect to the standard electroweak prediction is due to
the exchange of a new unmixed $Z'$.}
\end{figure}

\newpage

\def \ajp#1#2#3{Am.~J.~Phys.~{\bf#1}, #2 (#3)}
\def \apny#1#2#3{Ann.~Phys.~(N.Y.) {\bf#1}, #2 (#3)}
\def \app#1#2#3{Acta Phys.~Polonica {\bf#1}, #2 (#3)}
\def \arnps#1#2#3{Ann.~Rev.~Nucl.~Part.~Sci.~{\bf#1}, #2 (#3)}
\def \cmts#1#2#3{Comments on Nucl.~Part.~Phys.~{\bf#1}, #2 (#3)}
\def \cn{Collaboration}
\def \cp89{{\it CP Violation,} edited by C. Jarlskog (World Scientific,
Singapore, 1989)}
\def \dpfa{{\it The Albuquerque Meeting: DPF 94} (Division of Particles and
Fields Meeting, American Physical Society, Albuquerque, NM, Aug.~2--6, 1994),
ed. by S. Seidel (World Scientific, River Edge, NJ, 1995)}
\def \dpff{{\it The Fermilab Meeting: DPF 92} (Division of Particles and Fields
Meeting, American Physical Society, Batavia, IL., Nov.~11--14, 1992), ed. by
C. H. Albright \ite~(World Scientific, Singapore, 1993)}
\def \efi{Enrico Fermi Institute Report No. EFI}
\def \epjc#1#2#3{Eur.~Phys.~J.~C~{\bf #1}, #2 (#3)}
\def \epl#1#2#3{Europhys.~Lett.~{\bf #1}, #2 (#3)}
\def \f79{{\it Proceedings of the 1979 International Symposium on Lepton and
Photon Interactions at High Energies,} Fermilab, August 23-29, 1979, ed. by
T. B. W. Kirk and H. D. I. Abarbanel (Fermi National Accelerator Laboratory,
Batavia, IL, 1979}
\def \hb87{{\it Proceeding of the 1987 International Symposium on Lepton and
Photon Interactions at High Energies,} Hamburg, 1987, ed. by W. Bartel
and R. R\"uckl (Nucl.~Phys.~B, Proc.~Suppl., vol. 3) (North-Holland,
Amsterdam, 1988)}
\def \ib{{\it ibid.}~}
\def \ibj#1#2#3{~{\bf#1}, #2 (#3)}
\def \ichep72{{\it Proceedings of the XVI International Conference on High
Energy Physics}, Chicago and Batavia, Illinois, Sept. 6 -- 13, 1972,
edited by J. D. Jackson, A. Roberts, and R. Donaldson (Fermilab, Batavia,
IL, 1972)}
\def \ijmpa#1#2#3{Int.~J.~Mod.~Phys.~A {\bf#1}, #2 (#3)}
\def \ite{{\it et al.}}
\def \jpb#1#2#3{J.~Phys.~B {\bf#1}, #2 (#3)}
\def \lkl87{{\it Selected Topics in Electroweak Interactions} (Proceedings of
the Second Lake Louise Institute on New Frontiers in Particle Physics, 15 --
21 February, 1987), edited by J. M. Cameron \ite~(World Scientific, Singapore,
1987)}
\def \ky85{{\it Proceedings of the International Symposium on Lepton and
Photon Interactions at High Energy,} Kyoto, Aug.~19-24, 1985, edited by M.
Konuma and K. Takahashi (Kyoto Univ., Kyoto, 1985)}
\def \mpla#1#2#3{Mod.~Phys.~Lett.~A {\bf#1}, #2 (#3)}
\def \nc#1#2#3{Nuovo Cim.~{\bf#1}, #2 (#3)}
\def \np#1#2#3{Nucl.~Phys.~{\bf#1}, #2 (#3)}
\def \pisma#1#2#3#4{Pis'ma Zh.~Eksp.~Teor.~Fiz.~{\bf#1}, #2 (#3) [JETP Lett.
{\bf#1}, #4 (#3)]}
\def \pl#1#2#3{Phys.~Lett.~{\bf#1}, #2 (#3)}
\def \pla#1#2#3{Phys.~Lett.~A {\bf#1}, #2 (#3)}
\def \plb#1#2#3{Phys.~Lett.~B {\bf#1}, #2 (#3)}
\def \pr#1#2#3{Phys.~Rev.~{\bf#1}, #2 (#3)}
\def \prc#1#2#3{Phys.~Rev.~C {\bf#1}, #2 (#3)}
\def \prd#1#2#3{Phys.~Rev.~D {\bf#1}, #2 (#3)}
\def \prl#1#2#3{Phys.~Rev.~Lett.~{\bf#1}, #2 (#3)}
\def \prp#1#2#3{Phys.~Rep.~{\bf#1}, #2 (#3)}
\def \ptp#1#2#3{Prog.~Theor.~Phys.~{\bf#1}, #2 (#3)}
\def \ptps#1#2#3{Prog.~Theor.~Phys.~Suppl.~{\bf#1}, #2 (#3)}
\def \rmp#1#2#3{Rev.~Mod.~Phys.~{\bf#1}, #2 (#3)}
\def \sci#1#2#3{Science {\bf#1}, #2 (#3)}
\def \si90{25th International Conference on High Energy Physics, Singapore,
Aug. 2-8, 1990}
\def \slc87{{\it Proceedings of the Salt Lake City Meeting} (Division of
Particles and Fields, American Physical Society, Salt Lake City, Utah, 1987),
ed. by C. DeTar and J. S. Ball (World Scientific, Singapore, 1987)}
\def \slac89{{\it Proceedings of the XIVth International Symposium on
Lepton and Photon Interactions,} Stanford, California, 1989, edited by M.
Riordan (World Scientific, Singapore, 1990)}
\def \smass82{{\it Proceedings of the 1982 DPF Summer Study on Elementary
Particle Physics and Future Facilities}, Snowmass, Colorado, edited by R.
Donaldson, R. Gustafson, and F. Paige (World Scientific, Singapore, 1982)}
\def \smass90{{\it Research Directions for the Decade} (Proceedings of the
1990 Summer Study on High Energy Physics, June 25--July 13, Snowmass, Colorado),
edited by E. L. Berger (World Scientific, Singapore, 1992)}
\def \tasi90{{\it Testing the Standard Model} (Proceedings of the 1990
Theoretical Advanced Study Institute in Elementary Particle Physics, Boulder,
Colorado, 3--27 June, 1990), edited by M. Cveti\v{c} and P. Langacker
(World Scientific, Singapore, 1991)}
\def \yaf#1#2#3#4{Yad.~Fiz.~{\bf#1}, #2 (#3) [Sov. J. Nucl. Phys. {\bf #1},
#4 (#3)]}
\def \zhetf#1#2#3#4#5#6{Zh.~Eksp.~Teor.~Fiz.~{\bf #1}, #2 (#3) [Sov. Phys. -
JETP {\bf #4}, #5 (#6)]}
\def \zpc#1#2#3{Zeit.~Phys.~C {\bf#1}, #2 (#3)}
\def \zpd#1#2#3{Zeit.~Phys.~D {\bf#1}, #2 (#3)}

\end{document}